\documentclass[showpacs, oneside, twocolumn, prl, amsmath, amssymb, nofootinbib, superscriptaddress]{revtex4-1}
\pdfoutput=1
\usepackage{cases}
\usepackage{amsmath}
\usepackage{amssymb}
\usepackage{amsfonts}
\usepackage{amssymb}
\usepackage{dcolumn}
\usepackage{bm}
\usepackage{bbm}
\usepackage{graphicx}
\usepackage{xcolor}
\usepackage{array}
\usepackage{subfigure}
\usepackage{hyperref}
\usepackage{wasysym}

\newcommand{\be}{\begin{equation}}
\newcommand{\ee}{\end{equation}}
\newcommand{\ba}{\begin{eqnarray}}
\newcommand{\ea}{\end{eqnarray}}

\newcommand{\gsim}{\mathrel{\hbox{\rlap{\lower.55ex \hbox {$\sim$}}
                   \kern-.3em \raise.4ex \hbox{$>$}}}}
\newcommand{\lsim}{\mathrel{\hbox{\rlap{\lower.55ex \hbox {$\sim$}}
                   \kern-.3em \raise.4ex \hbox{$<$}}}}

\hypersetup{colorlinks=true,
            breaklinks=true,
            pdfstartview=Fit,
            linkcolor=blue,
            citecolor=blue,
            urlcolor=blue}

\begin{document}



\title{New test on general relativity and $f(T)$ torsional gravity from galaxy-galaxy weak lensing surveys}

\author{Zhaoting Chen}
\email{zhaoting.chen@postgrad.manchester.ac.uk}
\affiliation{Department of Astronomy, School of Physical Sciences, University of Science and Technology of China, Hefei, Anhui 230026, China}
\affiliation{CAS Key Laboratory for Research in Galaxies and Cosmology, University of Science and Technology of China, Hefei, Anhui 230026, China}
\affiliation{School of Astronomy and Space Science, University of Science and Technology of China, Hefei, Anhui 230026, China}
\affiliation{Jodrell Bank Centre for Astrophysics, Alan Turing Building, School of Physics and Astronomy, The University of Manchester, Oxford Road, Manchester, M13 9PL, United Kingdom}

\author{Wentao Luo}
\email{wentao.luo@ipmu.jp}
\affiliation{Department of Astronomy, School of Physical Sciences, University of Science and Technology of China, Hefei, Anhui 230026, China}
\affiliation{Institute for the Physics and Mathematics of the Universe (Kavli IPMU, WPI), UTIAS, Tokyo Institutes for Advanced Study, University of Tokyo, Chiba, 277-8583, Japan}

\author{Yi-Fu Cai}
\email{yifucai@ustc.edu.cn}
\affiliation{Department of Astronomy, School of Physical Sciences, University of Science and Technology of China, Hefei, Anhui 230026, China}
\affiliation{CAS Key Laboratory for Research in Galaxies and Cosmology, University of Science and Technology of China, Hefei, Anhui 230026, China}
\affiliation{School of Astronomy and Space Science, University of Science and Technology of China, Hefei, Anhui 230026, China}

\author{Emmanuel N. Saridakis}
\email{msaridak@phys.uoa.gr}
\affiliation{Department of Astronomy, School of Physical Sciences, University of Science and Technology of China, Hefei, Anhui 230026, China}
\affiliation{Department of Physics, National Technical University of Athens, Zografou Campus GR 157 73, Athens, Greece}

\begin{abstract}
We use galaxy-galaxy weak lensing data to perform a novel test on general relativity (GR) and $f(T)$ torsional gravity. In particular, we impose strong constraints using the torsional (teleparallel) formulation of gravity in which the deviation from GR is quantified by a single parameter $\alpha$, an approximation which is always valid at low-redshift Universe and weak gravitational fields. We calculate the difference in the deflection angle and eventually derive the modified excess surface density profile, which is mainly affected at small scales. Confronting the predictions with weak lensing data from the Sloan Digital Sky Survey Data Release 7, we obtain the upper bound on the deviation parameter, which, expressed via the dimensionless percentage in the Universe energy content, reads as ${\rm{log_{10}}}\Omega_\alpha\le -18.52_{-0.42}^{+0.80}$({stat})$_{-0.37}^{+1.50}$({ sys})$ [R_c/0.015R_{200}]$ with systematics mainly arising from the modelling of astrophysics, upon a reasonable choice of cut-off radius for
$f(T)$ gravity in form of $T + \alpha T^2$.
To our knowledge, this is the first time that GR is verified at such an accuracy at the corresponding scales.
\end{abstract}



\maketitle

\section{Introduction}

As was initiated by Einstein over a century ago, investigations on the relation between gravitation and geometry provide insightful views on the nature of spacetime. Nevertheless, slight modifications from general relativity (GR) have been proposed to explain cosmic acceleration at late and early times \cite{Capozziello:2011et}. Such deviations from GR can be formulated either in the usual curvature formalism \cite{Nojiri:2006ri}, or in the equivalent torsional (teleparallel) one \cite{Cai:2015emx}. Recently, it has become known that observations of the Large Scale Structure (LSS) of the Universe can be an effective probe to examine cosmological paradigms \cite{vanUitert:2017ieu, SpurioMancini:2019rxy}. Accordingly, it is suggested that a mild tension exists between the Cosmic Microwave Background (CMB) and LSS observations \cite{Douspis:2018xlj}, hinting towards the possibility of a new theoretical framework beyond $\Lambda$ Cold Dark Matter ($\Lambda$CDM) cosmology or even beyond GR. Hence, it is interesting to explore GR and its alternatives in the context of LSS probes.

Measurements of gravitational lensing are very efficient towards such a direction. For instance, Refs. \citep{Brouwer:2016dvq, Luo:2020iup} uses gravitational lensing to test emergent gravity, while Ref. \cite{Barreira:2015fpa} studies the lensing potential around void regions based on the assumption of cubic Galileon and Nonlocal gravity cosmologies. In particular, in this approach one typically
considers a usual Schwarzschild geometry in the framework of GR and modifies the excess gravitational potential by introducing mass components that yield additional impacts on the excess surface density (ESD) profile. Thus, it is natural to test theories of modified gravity using weak lensing measurements in various experimental environments, such as void lensing \cite{Baker:2018mnu}, cluster formation and its density profile \cite{Lombriser:2011zw}, cluster lensing \cite{Narikawa:2012tg, Narikawa:2013pjr}, etc. However, in most of pioneer studies, the effective lensing potential has been treated as the average of two scalar potentials from the metric perturbation following the geodesic equation. This fails to consider the geometric contribution to the local Euclidean definition of angle, which was first pointed out in Ref. \cite{Rindler:2007zz}, if one attempts to deal with no asymptotic flatness of Schwarzschild-de Sitter (SdS) spacetime.

In this paper, we use the new observational signatures on light-bending geometry to present an innovative test on GR, by using galaxy-galaxy weak lensing measurements. In particular, possible deviations from GR, quantified by the torsional formalism for convenience, alter the space-time geometry and thus affect the light-bending trajectory \cite{Ruggiero:2016iaq}. Hence, with such an implication on light-bending geometry we show that the ESD profile at small scales can be very sensitive to this effect, {and thus, put tight constraints on GR}. In order to illustrate the capability of this novel method, we make use of the galaxy group catalog and weak lensing shear catalog from the Sloan Digital Sky Survey (SDSS) Data Release (DR) 7 \cite{Abazajian:2008wr}, to {rule out a class of} deviations from GR.

\section{Formalisms}

We begin with a brief discussion on torsional formulation of gravity following Ref. \cite{Cai:2015emx}. First of all, in this framework the (equivalence of) GR Lagrangian reads as ${\cal{L}}_{GR} = T -2 \Lambda$, where $\Lambda$ is the cosmological constant and $T$ the torsion scalar that arises from contractions of the torsion tensor, similar to obtain the Ricci scalar in GR. {Given a homogeneous and isotropic universe described by} Friedmann-Lema\^{i}tre-Robertson-Walker (FLRW) metric, the torsion scalar becomes $T=-6H^2$, where $H$ is the Hubble function. {For a} spherically symmetry spacetime we acquire $T\propto r^{-2}$, where $r$ is the
radial coordinate. Hence, in the case of low-redshift Universe and weak gravitational fields that we focus in this study, any deviation from GR can be quantified as:
\begin{equation}
 f(T) = -2\Lambda+T+\alpha T^2+{\cal{O}}(T^3) ~,
\label{actionfull}
\end{equation}
where the parameter $\alpha$ has the dimension of length squared.

It has been shown that for the theory (\ref{actionfull}) there exists a general class of spherically symmetric solutions $ds^2 = c^2 e^{A(r)} dt^2 - e^{B(r)} dr^2 - r^2 d\Omega$ of the form \cite{Ruggiero:2016iaq, Farrugia:2016xcw}
\begin{align}
 A(r) &= -\frac{2GM}{c^2 r}-\frac{\Lambda}{3}r^2-\frac{32\alpha}{r^2}
\nonumber\\
 B(r) &= \frac{2GM}{c^2 r}+\frac{\Lambda}{3}r^2+\frac{96\alpha}{r^2} ~,
\label{solutionAB}
\end{align}
where we have neglected terms ${\cal{O}}\left(\alpha/r^2\right)^2$ due to the weak-field limit. As expected, $\alpha$ quantifies the deviation of the usual Schwarzschild-de Sitter solution of GR. With this gravitational potential, the deflection angle at non-cosmological scales reads
\cite{Ruggiero:2016iaq, Butcher:2016yrs}:
\begin{equation}
 \hat{\beta}=4GM/Rc^2+40\pi\alpha /R^2 ~,
\label{lightbending}
\end{equation}
where $R$ is the impact factor and $M$ is the mass of the point source.

It is worth mentioning that, Eq. \eqref{solutionAB} can be also used to test GR in the Solar System in terms of lensing, time delay and perihelion precession, where solar system generally serves as an ideal point mass as in Refs. \cite{Iorio:2012cm, Farrugia:2016xcw}. Such constraints on $\alpha$ can be converted into the relative correction to the standard Newtonian potential. 
However, solar system scales are too small to allow us to probe the connection between $\alpha$ and cosmic expansion, structure formation, etc., due to the fact the underlying theories are possibly scale dependent. In contrast, weak lensing provides a novel window, complimentary to the Solar System, to test those theories at relatively large scales. As shown in the following, weak lensing observables can be linked directly to corrections beyond Lambda Cold Dark Matter ($\Lambda$CDM) under a $f(T)$ gravity parametrization.

\section{Weak lensing effects}
The effective lensing potential $\psi(\vec{\xi})$ is defined as
\begin{equation}
 \psi(\vec{\xi})=\frac{D_{ds}}{D_d D_s}\frac{2}{c^2}\int \Phi(\vec{\xi},z)dz ~,
\end{equation}
where $\vec{\xi}$ is the position on the lens plane and $z$ is the comoving angular distance to lens plane, while $D_{ds}$, $D_d$, $D_s$ are, respectively, the angular distances between lens and source, lens and observer, and observer and source. In GR, this potential can be directly read from $A$. However, since in deviations from GR parametrized by \eqref{solutionAB} $A$ and $B$ are asymmetric function,
the lensing potential should be calculated via $\vec{\beta} = D_d \vec{\nabla}_{\xi} \psi$ (note that $\vec{\beta}= \frac{D_{ds}}{D_s} \vec{\hat{\beta}}$). Although this does not provide a unique solution, it is expected that the physical solution should be $\Phi(1/r)$ without a constant. Hence, it leads to
\begin{equation}
 \Phi(\vec{\xi},z)=\Phi_{\rm{Newton}}-20\frac{\alpha c^2}{r^2} ~,
\end{equation}
with $r=(\xi^2+z^2)^{1/2}$. Using the weak lensing formalism \cite{Schneider:2005ka}, we can calculate the convergence under the relation $\kappa(\xi) = D^2_d \nabla^2_\xi \psi/2$ as
\begin{equation}
 \kappa = \frac{4\pi G}{c^2} \frac{D_d D_{ds}}{D_s} \left[ \Sigma(R) - \frac{10 \alpha c^2}{G R^3} \right] ~,
\end{equation}
where $\Sigma$ denotes the surface mass density. It is natural to define an effective surface mass density $\Sigma_{eff}$:
\begin{equation}
 \Sigma_{\rm{eff}}=\Sigma-\frac{10\alpha c^2}{G R^3} ~.
\end{equation}
To acquire a more explicit picture of the physical meaning of this modification, we recall the relation between $\alpha$ and the critical density of the additional (torsional) energy component $\Omega_{\alpha}^{0}$ introduced by $\alpha T^2$, namely,  $\alpha=c^2 \Omega_{\alpha}^{0} / (18 H_0^2)$ \cite{Nesseris:2013jea, Basilakos:2018arq, Anagnostopoulos:2019miu}.

\begin{figure*}[!]
\includegraphics[width=0.6\textwidth]{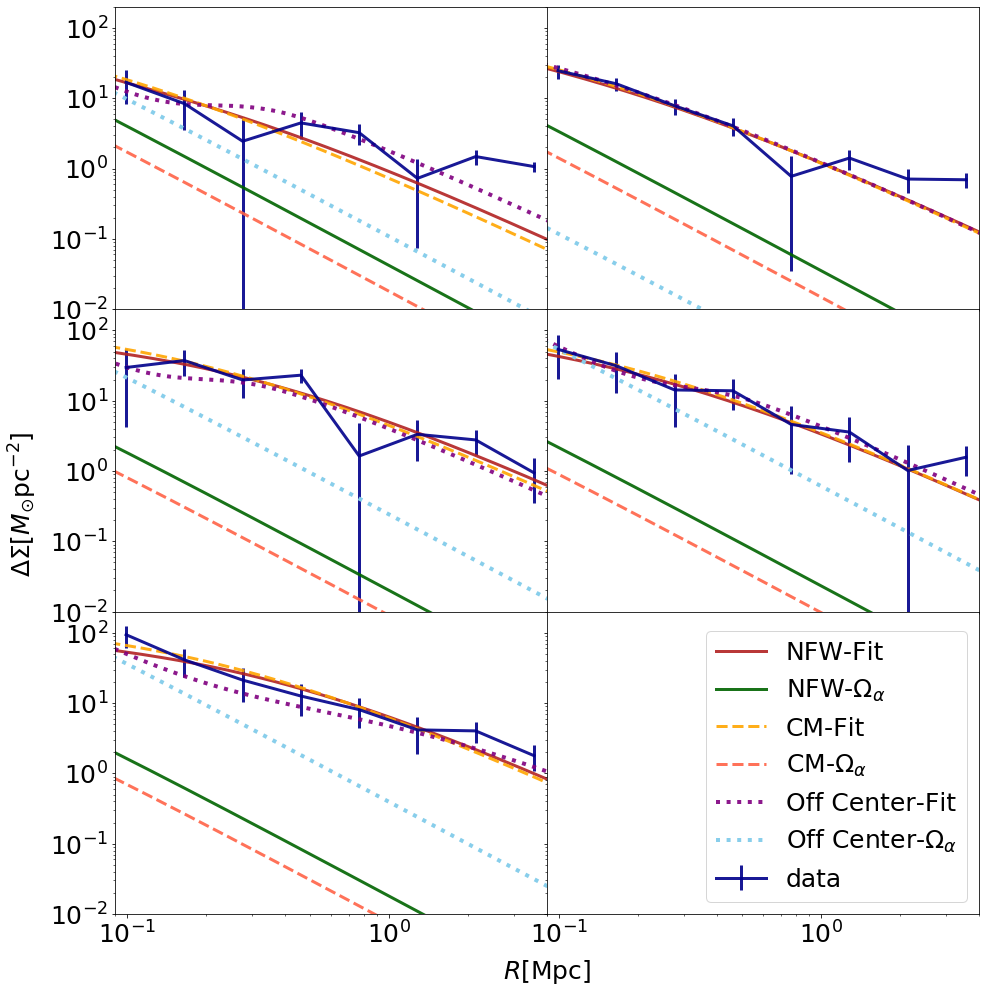}
\caption{
Best fits to ESD profile. The horizontal axis is the projected distance away from the lens galaxy, while the vertical axis corresponds to lensing signals from the component and total ESD profiles. The data points are galaxy-galaxy weak lensing measurements around spectroscopic galaxies within different stellar mass bins. The errors are estimated based on bootstrap sampling. The overall fitted ESD is denoted with ``-Fit'' and contribution from beyond GR extension is denoted with ``-$\Omega_\alpha$''. Three different models, simple NFW with free concentration parameters (``NFW-''), simple NFW with Gaussian prior for concentration parameter calculated according to the concentration-mass relation reported in Ref. \cite{Neto:2007vq} (``CM-''), and NFW with off-center effect (``Off Center-'') are shown in the legends.
}
\label{fig:ESD}
\end{figure*}

As discussed previously, in order to apply this modification, the scale of the projected radius far exceeds the scale that encloses most of stellar mass component. As a reasonable consequence it is obvious that the deviation term diverges when averaging from $0$ to the projected radius $R$, suggesting that a cut-off radius $R_c$ should be imposed. Here we primarily consider $R_c=R_{1/2}$, where $R_{1/2} \approx 0.015 R_{200}$ is the radius that encloses half of stellar mass (note that for a typical halo $M_{200}=10^{12} M_{\odot}$ at $z=0.1$, $R_c\sim4 ~ \rm{kpc}$), which has a universal linear relation to the virial radius of the galaxy \cite{Kravtsov:2012jn}.
We emphasize that this choice of $R_c$ is meant to be a conservative estimate of the scales where this effective description of $f(T)$ holds, and does not affect our methodology of constraining $\alpha$. As we show below, when $R_c$ is small, the effect of $\alpha$ is simply $\sim 1/R_c$, and therefore $\alpha/R_c$ can be constrained as one variable.

Defining $\epsilon \equiv R_c/R$, the modified ESD profile takes the form
\begin{equation}
 \Delta\Sigma_{\rm{eff}}(R) = \Delta\Sigma(R) + \frac{5 c^4 \Omega_{\alpha}^0 (2 - \epsilon - \epsilon^2)}{9 G H_0^2 R^3 \epsilon(1 +\epsilon)} ~.
\end{equation}
Hence, the effect of the induced modification due to the deviation from GR is
\begin{equation}
 \Delta\Sigma_{\alpha}(R) \approx 2.3 \frac{M_{\odot}}{\rm{pc^2}} \frac{\Omega_{\alpha}^0}{10^{-17}} \Big( \frac{1\rm{Mpc}}{R} \Big)^2 \frac{1 \rm{kpc}}{R_c} \Big( \frac{2 -\epsilon -\epsilon^2}{1 +\epsilon} \Big) ~.
\end{equation}
We consider this effect to be a modification to the regular $\Lambda$CDM halo, which can be modeled as a halo with Navarro-Frenk-White (NFW) density profile \cite{Navarro:1996gj}. Accordingly, for consistency we require $\alpha$, or equivalently $\Omega_{\alpha}^0$, to be small so that the halo formation is not affected \cite{Wu:2012hs, Finch:2018gkh, Cai:2019bdh, Bahamonde:2019zea}.

\section{Results}

Now we are ready to test GR and possible deviation with galaxy-galaxy weak lensing data. We use data selected from group catalog based on SDSS DR7 built in Refs. \cite{Yang:2007yr, Luo:2016ibp}. The samples are then subdivided into different stellar mass bins similar to Refs. \cite{Brouwer:2016dvq, Luo:2020iup}. We present detailed descriptions of this selection in the Appendix. The density profile can be modeled according to Ref. \cite{Wright:1999jc}, considering the off-center effect and stellar component.

Note that in principle there exist mechanisms that can modify the NFW ESD profile within GR, such as sub-halo contributions \cite{Mandelbaum:2004mq} to have similar effects on small scales. However, since we will demonstrate that $\Omega_\alpha$ is severely constrained by weak lensing, we consider the following three scenarios to extract the upper bound of the deviation parameter $\alpha$.
First, we use a simple NFW with $\alpha$ and treat halo mass and concentration as free parameters to demonstrate how an extremely small energy fraction of $\alpha$ can provide sizeable impact on small scales, as shown by the ``NFW-'' fit in Fig. \ref{fig:ESD}. Note that, due to the degeneracy between concentration and other effects mentioned above, the concentration parameters are suppressed. Second, we adopt the concentration-mass relation (``CM-'') proposed in Ref. \cite{Neto:2007vq} as a Gaussian prior
\footnote{For each chain, the Gaussian prior is dynamically changed so that mean of the prior is calculated via the CM relation. We fix the variance of the prior to be unity.}
to suppress the contribution from $\alpha$, which tests the lower limit of our desired upper bound as shown by the ``CM-'' fit in Fig. \ref{fig:ESD}. Third, as a very conservative estimation, we allow $\alpha$ to vary in different mass bins and adopt off-center effect (``Off Center-'') to further suppress the contribution from NFW halo on small scales, which tests the upper limit of our desired upper bound as shown by the ``Off Center-'' fit in Fig. \ref{fig:ESD}.

For each stellar mass bin, we set $\alpha$ to be an independent parameter for each fit, demonstrating the effect of $\Delta\Sigma_{\alpha}$. Since the typical $R_c$ is much smaller than the smallest scale in the plot (approximately $50$ kpc), the contribution coming from $\Delta\Sigma_{\alpha}$ is a power law of $R$. As the data indicate, additional effects on small scales should lift the ESD profile, suggesting a positive $\alpha$.
{For these five datasets, the fit for each scenario is roughly the same in terms of reduced $\chi^2$ statistics. As excepted, the ``CM-'' fit gives a low amplitude of $\alpha$, opposite to the ``Off Center-'' case. An intermediate fit of ``NFW-'' estimates a contribution of $\Delta\Sigma_\alpha\sim10M_{\odot}$pc$^{-2}$ at scales smaller than $0.1$ Mpc. This confirms that ESD on small scales is very sensitive to modified light bending beyond GR.}

Then we estimate the upper bound for $\alpha$. Following the argument above, we use the result of ``NFW-'' as our mean and statistical error. Theoretical systematics are estimated with ``CM-'' and ``Off Center-'', with ``CM-'' providing a lower limit and the fourth stellar mass bin of ``Off Center-'' fit providing an upper limit. In summary, the upper bound of $\alpha$ is given by,
\begin{equation}
 \alpha \le 0.33_{-0.21}^{+1.76} ~ {\rm pc}^2 ~ [\frac{R_c}{0.015 R_{200}}] ~, \nonumber
\end{equation}
or equivalently:
\begin{equation}
{\rm{log_{10}}}\Omega_\alpha\le -18.52_{-0.42}^{+0.80} ~ [\frac{R_c}{0.015 R_{200}}] ~,
\end{equation}
at 68\% confidence level with an estimation of systematic uncertainty of $_{-0.37}^{+1.50}$ from modelling.
Here we mention $R_c$ to remind that the effective constraint is scale dependent.
As we can see, the galaxy-galaxy weak lensing analysis implies that GR is verified to an order of $\sim10^{-18}$, and any possible deviation beyond GR on galaxy scales should satisfy this upper bound.

\begin{figure*}[!]
\includegraphics[width=0.6\textwidth]{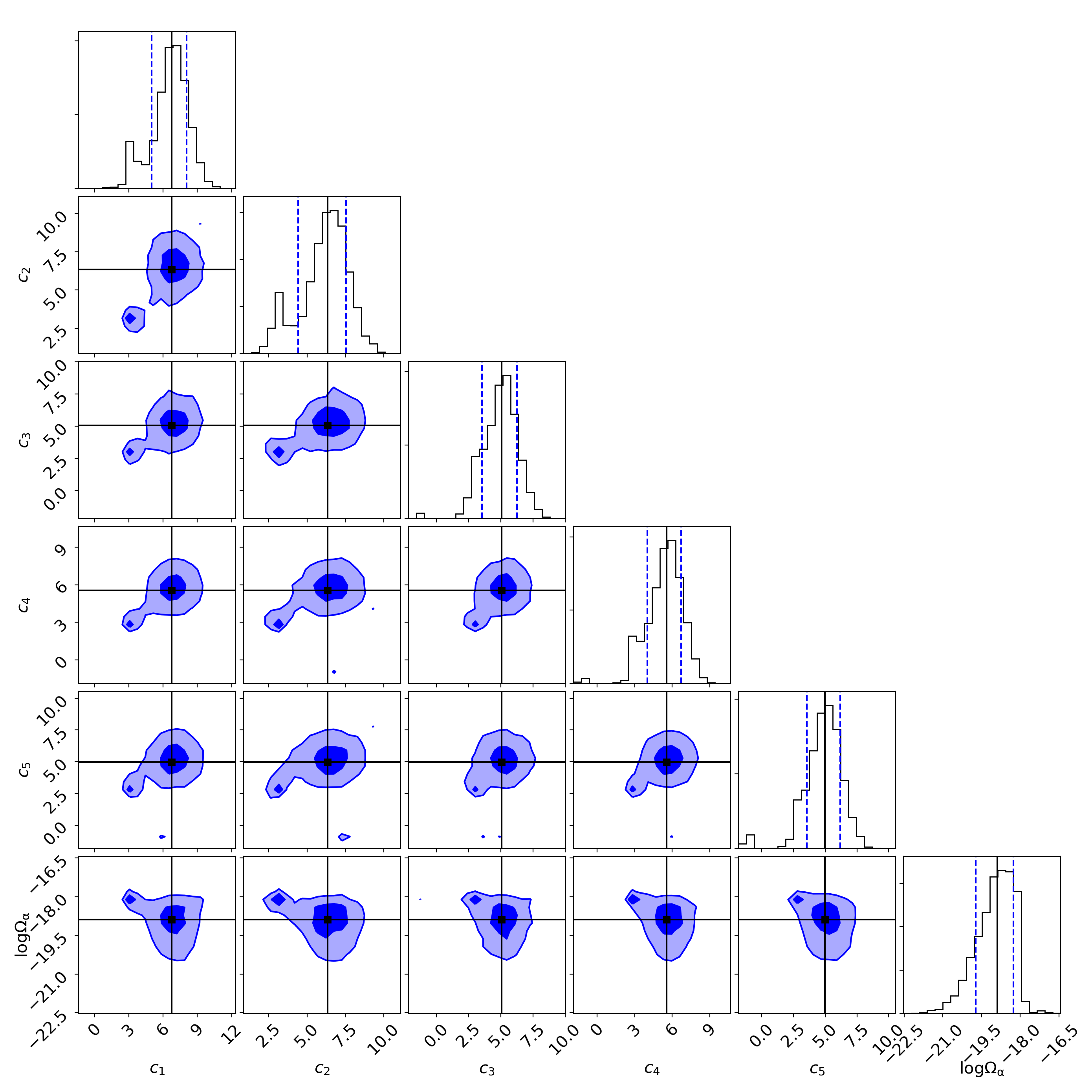}
\caption{
1$\sigma$ and 2$\sigma$ confidence regions from ``CM-Fit''.
}
\label{fig:contour}
\end{figure*}

Our result suggests that such modification to GR is extremely well constrained, as the energy density contribution could become very small. Therefore, such modification would yield a limited effect on structure formation. An intuitive way is to examine the modified Poisson equation for gravitational potential and matter overdensity at subhorizon in the quasistatic approximation: $k^2 \phi = 4\pi G_{\rm{eff}} a^2 \delta\rho$, in which the modification brought by beyond GR effect is $G_{\rm{eff}}/G = 1/(1+2\alpha T/c^2)$ \cite{Nunes:2018xbm, Yan:2019gbw}. With the above bound on $\alpha$, one gets $|1 - G_{\rm{eff}} / G | \sim 10^{-19}$. Thus our analysis should be consistent with the observed structure formation.
Recall that ``CM-Fit'' uses known concentration-mass relation as a prior. In Fig.~\ref{fig:contour}, we show the two-dimensional posteriors obtained by this fit. All the concentration parameters tend to have lower values if the model only apply a simple NFW density profile, as the weak lensing signal is a stacked measurement from halos with different off-center effect, triaxiality, satellite galaxies, and so on.

Because of the selection of the single galaxy system, the satellite galaxies are of least effects in our case. The off-center effect only suppresses signals at smaller scales, which are often tangled with the stellar contribution. It was shown in Ref. \cite{Luo:2017zbc} that this effect is roughly about $40~{\rm kpc/h}$ for our sample, and thus the off-center effect does not yield significant impact on our results either. The main concern may come from the fact that, SDSS DR7 spectroscopic sample has a detection limit of r band $17.77$, whereas the photometric catalog can reach up to $22.0$~magnitude. As a result, there may exist contaminations of stellar components of unidentified satellites for our galaxy sample. Recently, a study performed in Ref. \cite{Luo:2020iup}, which is based on the illustrisTNG300-3 simulation \cite{Nelson:2017cxy}, shows that the fraction of undetected satellite galaxies would contribute only $10\%$ to the maximum while including all the clusters in the most massive halo mass bin. However, a joint analysis in the future study is expected to quantify all these astrophysical effects accurately.

Note that the stacked profile somehow marginalizes the difference but smearing the inner profile due to the aforementioned effects. Yet, the degeneracy between concentration and $\alpha$ value pushes concentration to even smaller value with larger $\alpha$. This suppression of concentration from $\alpha$ is significant enough that it would lead to detectable observables which is contradicted by galaxy-galaxy lensing results in Ref. \cite{Luo:2017zbc}.

\section{Viability}

For a self-consistent check, we estimate the quantity $\alpha/r^2$. In particular, at length scales probed by weak lensing, it becomes $\alpha / r^2 \sim 10^{-8}$. This observation shows that we can indeed safely neglect the higher order terms from the solution \eqref{solutionAB} and the subsequent analyses.

As we made several assumptions and approximations, it is important to discuss the viability of our results by examining these impacts. For one thing, the aforementioned choice of the cut-off radius $R_c$ is conservative, and guarantees the robustness of our results. Note that the light bending geometry is based on two assumptions, i.e., spherical symmetry and weak field. We also study the case when $R_c$ breaks either the symmetry or the weak field limit.  Currently, we provide $\alpha$ at pc$^2$ level and the value of $\alpha/R_c^2$ is very small. If we take $\alpha/R_c^2 \sim1$, then the characteristic scale of $R_c$ will be much smaller and the upper bound of $\rm{log_{10}}\Omega_{\alpha}$ dramatically reduces to about $-20$, 2 order of magnitude tighter than our claim of an upper bound on $\rm{log_{10}}\Omega_{\alpha} \sim -18$. It is reasonable and responsible to take latter value as the upper bound.

Moreover, we further comment on the halo modeling effects. Here, we consider a simple model to demonstrate the effect. However, our results still hold in a robust way, that even if all the ESD signals at low $R$ are produced by $\alpha$, such as the middle row of graphs of Fig.~\ref{fig:ESD}, then the upper bound of $\rm{log_{10}}\Omega_{\alpha}$ can be uplift by one order, which still remains stringent. Additionally, uncertainties arising from the ESD signal reconstruction at small separations can yield a significant impact \cite{Leauthaud:2016jdb, McClintock:2018bxh}. Nevertheless, unless the contribution from $\alpha$ is 10 times higher or lower than the present case, our results remain viable.

\section{Conclusion}

In this paper, we used galaxy-galaxy weak lensing data to test GR and the possible departure. To demonstrate the strong constraints upon deviations from GR, we considered in particular the torsional formulation of gravity by introducing a new deviation parameter $\alpha$. By calculating the deflection angle in this modified gravity, we can derive the modified ESD profile, which is mainly affected at small scales. Therefore, confronting the theoretical predictions with weak lensing data from group catalog based on SDSS DR7 \cite{Yang:2007yr, Luo:2016ibp}, we are able to extract an upper bound on $\alpha$, which, expressed via the dimensionless percentage in the Universe energy content, reads as $\log_{10}{\Omega_\alpha^0} \le -18.52_{-0.42}^{+0.80} ~ [R_c/0.015R_{200}]$. To our knowledge, this is for the first time that GR has been verified at such an accuracy at the aforementioned scales. Moreover, a systematic error of $_{-0.37}^{+1.50}$ is roughly estimated upon the modeling of several astrophysical effects. It is noted that a joint analysis including all possible astrophysical effects should further improve this error, which deserves to be performed in the future study.
Our method provides a direct application for future surveys including Vera Rubin telescope \cite{Abell:2009aa}, Euclid \cite{Wallner:2017}, and WFIRST \cite{WFIRST:2019}. Especially, the Vera Rubin telescope with 20000 deg$^2$ (half of the sky) survey region and a depth of r band $26$ magnitude enables subpercent-level statistical errors in weak lensing measurements.

We conclude by highlighting the implications of the reported probe that can initiate several follow-up studies. We performed a novel weak galaxy-galaxy lensing test on GR. It is interesting to perform similar analyses to constraint cosmological scenarios beyond $\Lambda$CDM \cite{Zhang:2018mlj} or modifications such as $f(R)$ gravity \cite{Li:2017xdi}. As shown in this paper, however, for extensions beyond GR which are generically valid at low redshifts and in the weak field limit, quantified by a single parameter, we extracted very strict constraints. Phenomenologically, a crucial outcome is that, even for infrared modifications of GR having cosmological implications at large scales, their Taylor expansions naturally generate high curvature/torsion terms that can be strongly bounded by the results reported in this paper. Thus, our analyses shed light on the motivation of theoretical investigations on possible modifications to GR from fundamental theories yielding small deviations from GR. Moreover, in the era of precision astronomy, a combination of various observational windows can impose tighter and tighter probes to GR and alternative theories \cite{Cai:2018rzd, Li:2018ixg, Li:2019lsm}.

\section*{Acknowledgments}

We are grateful to T. Qiu, M. Sasaki, Y. Wang, P. Zhang and J. Zhu for valuable comments. This work is supported in part by the NSFC (No. 11722327, 11653002, 11961131007, 11421303, J1310021), by CAST Young Elite Scientists Sponsorship Program (2016QNRC001), by the National Youth Talents Program of China, and by the Fundamental Research Funds for the Central Universities. Z. C. wishes to thank the Kavli IPMU for kind hospitality. W. L. acknowledges the support from World Premier International Research Center Initiative (WPI) Japan. All numerics are operated on the computer clusters {\it Linda \& Judy} in the particle cosmology group at USTC.



\section*{Properties of selected lenses, sources and halos}

\subsection*{Lenses}

The lenses are from group catalog based on SDSS DR7 \cite{Yang:2007yr}, namely, $472419$ groups out of which $368020$ are with halo mass estimation based on abundance matching. To minimize the effects of nearby structures, we only select single galaxy systems which further reduce the number to $326172$. All the sizes of those single galaxy systems are quantified by virial radius, which is the radius containing $180$ overdensity compared to the mean density of the universe. There are 36,759 galaxies which have no spectroscopic redshift due to fiber collision, yet assigned to the redshift of their nearby galaxies. Those galaxies are included in the group-finding procedure, and thus the single systems are not effected by the fiber collision effect.

The samples are sub-divided into different stellar mass bins, similarly as in Ref. \cite{Brouwer:2016dvq}, while increasing the number of samples and including a higher stellar mass bin. The stellar mass of this sample is estimated based on Ref. \cite{Bell:2003cj}. The basic statistical properties of the binning of the sample are given in Table \ref{tab:tbl-1}.

\begin{table}[h!]
\begin{center}
\begin{tabular}{ccccc}
\hline
\\
$\log_{10} M_{st}$ range &$\ \ $ $N_{sat}$ $\ \ $ & $\langle z \rangle $ & $\langle
\log_{10} M_{st} \rangle$ &
$\langle \log_{10} M_h \rangle$ \\
\\
\hline
\\
  8.5-10.5    & 145 298 & 0.091 & 10.266 & 11.995 \\
  10.5-10.8   & 104 773 & 0.123 & 10.648 & 12.441 \\
  10.8-10.9   & 28 833  & 0.143 & 10.848 & 12.748 \\
  10.9-11.0   & 22 427  & 0.155 & 10.946 & 12.922 \\
  11.0-11.8   & 24 841  & 0.165 & 11.087 & 13.237 \\
\hline
\end{tabular}
\caption{\label{tab:tbl-1} Properties of the lens samples created for this paper.}
\end{center}
\end{table}

\subsection*{Sources}
The source catalog based on SDSS DR7 is from Ref. \cite{Luo:2016ibp}. The ESD is related to the stacked shape by the geometry factor $\Sigma_{cr}$ through $\gamma_T(R) \Sigma_{cr} = \Delta\Sigma(R)$. Instead of measuring $\gamma_T$, it is more convenient to directly measure $\Delta\Sigma(R)$ as a whole by applying both photometric and spectroscopic redshift so that
\begin{equation}
 \Delta\Sigma(R) = \frac{1}{2\mathcal{R}} \frac{\sum_{l,s} w_{l,s} e_{t,ls} [\Sigma_{cr,ls}^{-1}]^{-1}} {\sum_{l,s}w_{l,s}} ~.
\end{equation}
$\mathcal{R}$ is the responsivity of shear given a shape estimator, and we use a universal shear responsivity by using the shape distribution of the source sample that $\mathcal{R} = 1 -\sum_{l,s} e_{rms}^2 w_{l,s} / \sum_{l,s}w_{l,s}$. The weight contains not only shape noise and measurement error, but additionally the geometry factor $\Sigma_{cr}$, namely $w_{l,s} = (\Sigma_{cr,ls}^{-1})^2 / (\sigma_e^2 +e_{rms}^2)$, where $\sigma_e$ is the shape noise and $e_{rms}$ is the error caused by the sky background noise and Poisson noise of each galaxy. Finally, $l,s$ denotes each lens-source pair system.

The major systematics arise from photometric redshift error \cite{Hirata:2003cv, Mandelbaum:2005wv}, which can lead up to $3\%$ systematic errors in lensing measurements. Moreover, we apply the boost factor to account for the other contamination caused by photometric error, which leads to misidentification of low-$z$ galaxies as high-$z$ galaxies. The boost factor is actually the ratio between the number of galaxies within radius for the lens sample and random points of a survey $B(r) = n(R) / n_{rand}(R)$, and therefore the final measured ESD is multiplied by this factor.

\subsection*{NFW halo}
The ESD of a dark matter halo is typically modelled as an NFW halo \cite{Wright:1999jc} with off-center effect. The NFW halo profile consists of two parameters, the characteristic mass scale $M_{200}$ and the concentration $c$. We adopt the following concentration-mass relation to reduce the degree of freedom \cite{Neto:2007vq}: $c = 4.67 (M_{200} / 10^{14} h^{-1} M_{\odot})^{-0.11}$. The total ESD is a sum of several component,
\begin{equation}
 \Delta\Sigma(R) = \Delta \Sigma_{\rm{host}} + \Delta \Sigma_{\rm{sub}} + \Delta \Sigma_{*} + \Delta \Sigma_{2\rm{h}} ~,
\end{equation}
where host halo, satellite halo, stars and two-halo component are included. For a single off-center radius $R_{\rm{off}}$, the ESD profile changes to \cite{Johnston:2005uu}
\begin{equation}
 \Sigma(R|R_{\rm{off}}) = \frac{1}{2\pi} \int^{2\pi}_{0} \Sigma_{\rm{NFW}} (\sqrt{R^2 \! + \! R^2_{\rm{off}} \! + \! 2R_{\rm{off}} R\cos\theta}) d\theta ~,
\end{equation}
while the actual ESD profile of the host halo is the convolution between the off-center radius and $\Sigma(R|R_{\rm{off}})$, i.e., $\Sigma_{\rm{host}} = \int dR_{\rm{off}} P(R_{\rm{off}}) \Sigma(R|R_{\rm{off}})$, where $P(R_{\rm{off}}) = \rm{exp} [-(R_{\rm{off}}/R_{\rm{sig}})^2/2] R_{\rm{off}} / R^2_{\rm{sig}}$. For $\Delta\Sigma_{*}$, the stellar mass component can be simply treated as a point-mass \cite{George:2012xd}, namely, $\Delta\Sigma_{*}(R) = M_*/(2\pi R^2)$, where $M_*$ is the stellar mass of candidate central galaxy.

The two-halo term can be calculated through the halo-matter correlation function \cite{Seljak:2004ni}. As shown in Ref. \cite{Luo:2017zbc}, the two-halo term remains trivially small below the scale of $1$ Mpc and becomes significant at larger scales. Thus, for the first two stellar mass bins used in our data, we have neglected the largest R when fitting the ESD since at that scale the measured $\Delta\Sigma$ is below $1~M_{\odot}/\rm{pc}^2$.

There is some possibility that the candidate lens galaxy is not the true central galaxy and may contain the subhalo component. In addition, in the group finder, there are some possibilities that the central galaxy is an interloper and may contain its original host halo component. Hence, usually a subhalo component $\Delta \Sigma_{\rm{sub}}$ will be included. However, as this component also affects mainly the small scales, introducing it leads to degeneracy between $\Delta \Sigma_{\alpha}$ and $\Delta \Sigma_{\rm{sub}}$. Thus, we neglect this term for our modeling, and we deduce that this estimation is the upper bound of $\alpha$.

\end{document}